\documentclass[journal, 10pt]{IEEEtran}
\IEEEoverridecommandlockouts

\usepackage{amsmath}
\usepackage{mathtools}
\usepackage{cuted}
\usepackage{revsymb}
\usepackage[tight,footnotesize]{subfigure}
\usepackage{balance}
\usepackage{xcolor}
\usepackage{eqnarray}
\usepackage{commath}
\usepackage{graphicx}
\usepackage{textcomp}
\DeclareGraphicsExtensions{.eps}
\graphicspath{{./Figures/}}
\usepackage{multicol}
\usepackage{algorithm,algorithmic}
\usepackage{subfiles}
\usepackage{mathcomp}
\usepackage{subfig}
\usepackage{cite}
\usepackage{cases}
\usepackage{amssymb}

\usepackage{verbatim}
\normalsize

\usepackage[normalem]{ulem}

\begin{document}

\title{ Terahertz Induced Protein Interactions  in  a Random Medium }

 \author{Hadeel Elayan, Andrew W. Eckford, and Raviraj Adve
%
%
%

\thanks{We would like to acknowledge the support of the National Science and Engineering Research Council, Canada, through its Discovery Grant program.} 

\thanks{H. Elayan and R. Adve are with the Edward S. Rogers Department of Electrical and Computer Engineering, University of Toronto, Ontario, Canada, M5S 3G4 (e-mail: hadeel.mohammad@mail.utoronto.ca; rsadve@ece.utoronto.ca).}
\thanks{ A. Eckford is  with the Department of Electrical Engineering and Computer Science, York University, Ontario, Canada, M3J 1P3 (e-mail: aeckford@yorku.ca).}

\thanks{ This work has been presented in part in~\cite{key}.}}

\maketitle
\begin{abstract}  
 Folding of proteins into their correct native structure is key to their function. Simultaneously, the intricate interplay between cell movement and protein conformation highlights the complex nature of cellular processes. In this work, we demonstrate the impact of Terahertz (THz) signaling on controlling protein conformational changes in a random medium. Our system of interest consists of a communication link that involves a nanoantenna transmitter, a protein receiver, and a channel composed of moving red blood cells. Due to the system dynamics, we investigate the influence of both the fast and slow channel variations on protein folding. Specifically, we analyze the system's selectivity to asses the effectiveness of the induced THz interaction in targeting a specific group of proteins under fading conditions. By optimizing the selectivity metric with respect to the nanoantenna power and frequency, it is possible to enhance the controllability of protein interactions. Our  probabilistic analysis provides a new perspective regarding electromagnetically triggered protein molecules, their micro-environment and their interaction with surrounding particles. It helps elucidate how external conditions impact the protein folding kinetics and pathways. This results in not only understanding  the mechanisms underlying THz-induced protein interactions but also engineering these still-emerging tools.
\end{abstract}

\begin{IEEEkeywords}
Terahertz signals, protein interactions, probability of folding, fast fading, slow fading, selectivity, outage.
\end{IEEEkeywords}

 \section{Introduction}

 Over the past two decades, the field of nanotechnology has experienced significant advancements, prompting researchers to explore methods for establishing reliable communication between nanomachines. This communication enables nanosensors to independently transmit their sensing data, receive instructions from a central command center, and collaborate with other nanomachines when necessary~\cite{akyildiz2010internet}. To achieve nanoscale communication, scientists have put forward various solutions that consider both molecular and electromagnetic (EM) communication approaches. From the EM perspective, plasmonic nano-lasers, plasmonic nanoantennas, and single-photon detectors have all indicated the significance of the Terahertz (THz) band, which spans from 0.1 to 10 THz, as a crucial facilitator of communication at the nanoscale~\cite{jornet2013graphene}.

The viability of utilizing THz intra-body communication has been strengthened through the characterization of tissues in the THz frequency range~\cite{yang2015numerical}. As a result, scientists have focused on creating models that accurately capture signal propagation, photo-thermal effects, and sources of noise~\cite{elayan2017terahertz,indrawijaya2018simulation}. For example, the authors in~\cite{abbasi2016terahertz} presented a channel model for THz propagation inside the human skin. To verify the achieved results, the authors have compared their findings with THz time-domain spectroscopy measurements of a skin sample. Moreover, researchers have proposed mechanisms to overcome the high path-loss and enable wireless communication with deeper implants for nano-biosensing applications~\cite{sangwan2018increasing}. Recently, the authors in~\cite{jornet2023nanonetworking} presented an in-depth overview covering the state of the art in EM nanoscale communication and networking, where they discussed nanonetworking in the THz band and beyond while focusing on applications ranging from nano-bio interfaces to quantum communications.

THz radiation operates within the energy range of molecular interactions, which encompass hydrogen bonds, intermolecular forces, and the vibration of macromolecules~\cite{wei2018application}. This alignment of energy levels elucidates why biomolecular interactions are highly responsive to emerging THz techniques. Specifically, proteins, the fundamental components of cells, exhibit distinctive collective vibrational patterns in the THz frequency range. These patterns correspond to crucial functional modes, which have been identified as potential drivers of conformational alterations, ligand binding, and changes in oxidation state~\cite{markelz2008terahertz}.

The dynamics and conformational changes of proteins involve a multitude of motions driven by various mechanisms, occurring at different timescales and amplitudes. Protein motions involve a broad range of timeframes, from fast bond vibrations in the femtosecond range to slow and extensive domain movements in the millisecond range~\cite{stadler2016picosecond}. These variations have posed challenges in identifying universal principles governing protein folding~\cite{eaton1996fast}. Proteins also experience constant external stresses and crowding from other biomolecules within cells, membranes, or extracellular spaces~\cite{dave2015fast}.

In our previous work, we proposed a hybrid communication model that combines EM and molecular approaches by utilizing proteins as molecular machines within the human body~\cite{elayan2022toward}. We employed a Markov model to derive the mutual information and calculate the capacity of the communication system under different constraints. This analysis was conducted for both a two-state and a multi-state protein model~\cite{elayan2020information}. In this context, capacity serves as a quantitative measure that provides researchers with insights into the amount of information gained by the protein signaling pathway as the protein transitions from an unfolded to a folded state. 

In the aforementioned works, we treated the external nanoantenna force as a deterministic component, in which the incident EM field directly impacts the protein population. However, we have not considered the impact of channel randomness on the EM wave that triggers the desired protein. In a more realistic model, cells act as obstacles in the path between the transmitter (Tx) and receiver (Rx). In addition, rapid fluctuations in received signal strength occur over both short distances and short time periods due to the mobility of these particles. For instance, in a blood medium, red blood cells (RBCs) interfere with propagation in a time-varying manner. This phenomenon is similar to fading in wireless communications, with the impinging signal suffering from shadowing. The fading can be either slow or fast, depending on the time constant of the protein dynamics~\cite{key}.

To bridge the gap, we consider a communication link that consists of a nanoantenna Tx, a protein Rx, and a channel composed of RBCs. Since a large number of moving molecules exist between the communicating endpoints, the channel experiences fading, which randomly attenuates the transmitted nanoantenna signal. Therefore, we will examine the impact of both fast and slow fading on our ability to stimulate controlled protein folding behaviors. To evaluate the system's performance under fading conditions, we study the selectivity of the system, which was initially investigated in~\cite{elayan2022selectivity} to demonstrate the efficiency of the stimulated THz interaction for a protein population.

The proposed selectivity metric assesses the capability of the nanoantenna to induce a conformational change within the desired protein population without affecting untargeted proteins in the network. It requires suppressing the effect of undesired proteins and enhancing only the impact of targeted ones. Specifically, in this work, we explore the effect of fading on the ability to achieve selective protein interactions. The unpredictability of fading presents a challenge, but by optimizing the selectivity metric with respect to the nanoantenna power and frequency, it is possible to enhance the controllability of protein interactions. From this perspective, we make the following  contributions:\begin{itemize}
\item 
We determine the best fit of the  nanoantenna  force realizations based on the developed propagation model.
\item We study the phenomenon of  fast and slow fading in THz intra-body communication and its impact on folding dynamics. 
\item We formulate a joint optimization problem to retrieve the optimal parameters that maximize the selectivity in fading scenarios.
\item We introduce \textit{selectivity outage} as a metric that indicates the performance of the slow fading channel.
\end{itemize}

Studying how the blood channel affects protein dynamics offers a broader perspective on the functioning and interactions of proteins within a living organism. This approach helps bridge the gap between in-vitro experiments and in-vivo conditions. These insights could impact the way experiments are designed and controlled, as researchers might explore the channel as a tool to influence folding patterns, investigate specific aspects of folding mechanisms, or validate hypotheses related to protein folding processes.

The rest of the paper is organized as follows. In Sec.~\ref{Sec:Sec2}, we present the system model for the  stimulated protein dynamics. In Sec.~\ref{Sec:Sec3}, we present a propagation model to compute the power received by the protein and calculate the path-loss  experienced due to the shadowing imposed by the  particles in the channel. In Sec.~\ref{Sec:Sec4}, we formulate the  expression of  the steady-state energy absorbed by the protein subject to the random nanoantenna force and use it to modify the Boltzmann distribution. In Sec.~\ref{Sec:Sec5}, we introduce the concept of fading in an intra-body environment and examine its impact on the capability to invoke controlled interactions in a protein network. In Sec.~\ref{Sec:Sec6}, we  demonstrate our numerical results. Finally, we draw our conclusions in Sec.~\ref{Sec:Sec7}.

\section{System Model}
\label{Sec:Sec2}

Our system model involves a nanoantenna Tx and a protein Rx. The channel we focus on is a blood vessel that contains various intra-body particles. Among these particles, RBCs are the largest, with a size of approximately 7 microns, and they are also the most abundant, constituting 45$\%$ of the particles in human blood. Therefore, RBCs play a crucial role in governing the propagation of THz waves in the bloodstream. Additionally, the blood cells are surrounded by blood plasma, which mainly consists of water (92$\%$). We model the plasma as a lossy medium, characterized by the complex permittivity of water at THz frequencies~\cite{johari2017nanoscale}. 

It is noteworthy that in our current work, we consider a dipole nanoantenna designed for the incident frequency being used, which is tuned to the vibrational frequency of the desired protein, without delving into the specific details of the antenna design. For more information on the design of nanoantennas, we refer interested readers to the works in~\cite{jornet2013graphene} and~\cite{sangwan2021beamforming}.

Cells have receptor proteins that bind to signaling molecules and initiate physiological responses. In our model, we consider our protein as a G-protein-coupled receptor located on the cell's exterior. When the receptor protein receives the nanoantenna signal, it undergoes a conformational change.

\begin{figure}[h!]
\centering
\includegraphics[width=0.36\textwidth]{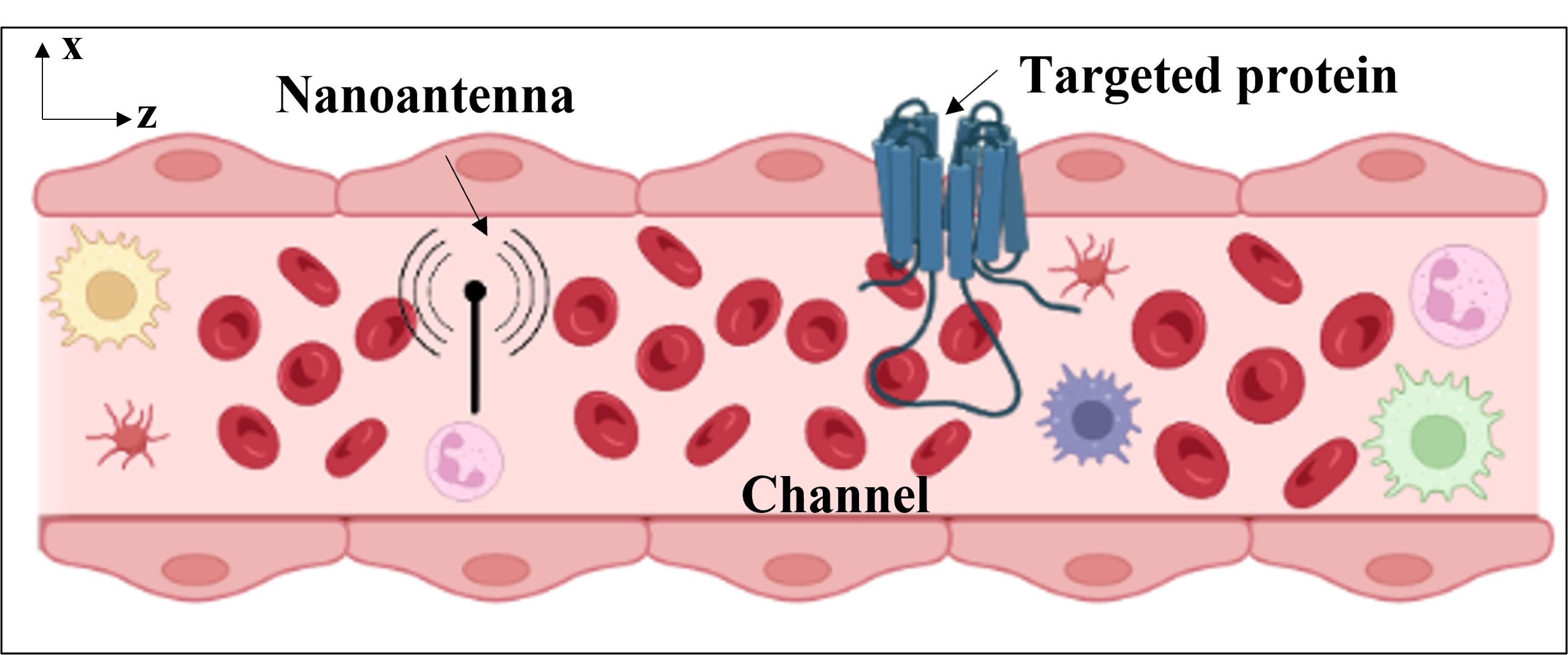}
\footnotesize
\caption{System model depicting a communication link: a nanoantenna Tx, a protein Rx and a channel composed of  particles. The model is created  with BioRender.com. }
\label{fig:system_model}
\end{figure}

Consider the blood vessel as depicted in Fig.~\ref{fig:system_model}. For an  $\mathbf{\hat x}$ polarized plane wave  propagating in the $\mathbf{+\hat z}$ direction, the electric field is given as~\cite{cheng1989field}
\begin{equation}
\begin{split}
\mathbf{E}(z,t)&=\mathbf{Re}\left[\mathbf{E}(z)e^{-j\omega t} \right] \mathbf{\hat x}\\
&=E_oe^{-\kappa z}\mathbf{Re}\left[{e^{j(\omega t-\bar \beta z)}}\right] \mathbf{\hat x}\\
&=E_o e^{-\kappa z} \cos(\omega t-\bar \beta  z) \mathbf{\hat x,}
\label{eq:electric_field}
\end{split}
\end{equation}
where $\mathbf{ E}({z})=E_o e^{-\varphi z}\mathbf{\hat x}$ is the linearly polarized electric field vector  phasor, $\varphi=\kappa+j\bar \beta$ is the complex wave number, $z$ is the propagation distance between the nanoantenna and the protein,  $\kappa$ is the attenuation constant (Nepers/m), $\bar \beta$ is the phase propagation constant (rad/m). Since the human body is a lossy medium, $\kappa$ and $\bar \beta$  have the following general forms~\cite{cheng1989field}
\begin{equation}
\kappa=\omega\sqrt{\left[ \frac{\ddot \mu\varepsilon'_{m}}{2}\left[ \sqrt{1+\left( \frac{\varepsilon''_{m}}{\varepsilon'_{m}} \right)^2} -1\right] \right]},
\end{equation}

\begin{equation}
\bar\beta=\omega\sqrt{\left[ \frac{\ddot\mu \varepsilon'_{m}}{2}\left[ \sqrt{1+\left( \frac{\varepsilon''_{m}}{\varepsilon'_{m}} \right)^2} +1\right] \right]}, 
\end{equation}
where $\omega=2\pi f$ is the angular frequency, $\varepsilon'_{m}$ and $\varepsilon''_{m}$ are the real and imaginary parts, respectively, of the complex permittivity of the medium ($\varepsilon^{*}_m=\varepsilon'_m-j\varepsilon''_m$), which we consider  as a combination of plasma  and RBCs, respectively; and $\ddot\mu$ is the relative permeability that acquires a value of one since most organic materials are non-magnetic at THz frequencies~\cite{lee2009principles}.

When a polarized object is subjected to the electric field given in~\eqref{eq:electric_field}, it induces a dipole moment. If the electric field is inhomogeneous, the field strength, and therefore the force, acting on each side of the particle will differ. Dielectrophoresis (DEP) is the force that arises from the interaction of a dielectric particle with the spatial gradient of an electric field. In our system, we model the protein as a sphere, which is a common approach for describing the DEP force acting on a particle~\cite{kim2018dielectrophoresis}.

The strength of the induced force, exerted on each end of the dipole, depends on both the medium and the protein's electrical properties, on the protein's shape and size, as well as on the frequency of the electric field. The time-averaged force acting on the protein is therefore given as~\cite{kim2018dielectrophoresis}
\begin{equation}
f_{o}=2\pi r^{3} \varepsilon_m \mathbf{Re}\left\{\mathcal{K} \right\}\nabla  |\mathbf{E}_{rms}|^2,
\label{eq:force_eq}
\end{equation}
where $\mathbf{E}_{rms}$ is the root mean square  value of the electric field given in~\eqref{eq:electric_field} and $|\mathbf{E}_{rms}|^2=\frac{1}{2}|E^2_o|e^{-2\kappa z}$.  In addition, $\nabla$ is the mathematical \textit{ Del vector operator}~($\nabla=\frac{\partial}{\partial x}\hat x+\frac{\partial}{\partial y}\hat y+\frac{\partial}{\partial z}\hat z$), and $\nabla  |\mathbf{E}_{rms}|^2=-\kappa |E^2_o|e^{-2\kappa z}$. From~\eqref{eq:force_eq}, $r$ is the protein radius, $\varepsilon_m=\varepsilon_{o}\varepsilon^*_m$ is the absolute permittivity of the medium, and $\mathcal{K}$ is the frequency-dependent polarization coefficient given by\begin{equation}
\mathcal{K}=\frac{\varepsilon^*_p-\varepsilon^*_m}{\varepsilon^*_p+2\varepsilon^*_m},
\end{equation}where $\varepsilon^{*}_p$ is the complex permittivity of the protein sample~\cite{knab2006hydration}.

\section{Propagation Model }
 \label{Sec:Sec3}
When an object is illuminated by a wave, the incident power experiences both absorption and scattering effects caused by particles. These phenomena contribute to power losses compared to the line-of-sight (LOS) component of the channel. In our specific propagation scenario, we utilize a first-order multiple scattering model to describe the propagation of THz waves~\cite{ishimaru1978wave}. This model accounts for the absorption and scattering that occur along the ray path. It is particularly applicable when particles primarily exhibit absorption rather than scattering behavior. This situation arises when the particle scattering albedo, which represents the ratio of scattering to total attenuation, is less than 0.5~\cite{enjamio2007rain}.

\subsection{Signal Transmission}
\label{Sec:Sec3A}
In our scenario, the nanoantenna transmits power $P_t$ through a distance $z$.  Since the THz beams are narrow, the beam pattern can be approximated by a Gaussian function. As such, the nanoantenna beam pattern is given by~\cite{holt1993development}
\begin{equation}
G_{t}(\theta_{t})=\exp\left[-4\ln 2\left( \frac{\theta_{t}}{\theta_{b}} \right)^2 \right],
\end{equation} where $\theta_{t}$ is the angle measured from the horizontal  direction, and $\theta_{b}$ is the 3 dB beamwidth angle. The Rx in our system is a protein. We treat it as a point Rx with a unity gain, i.e., $G_r(\theta_r)=1$. The received power is therefore given as
\begin{equation}
P_{r}=\frac{P_{t}\lambda_{g}^2 G_t(\theta_{t})G_r(\theta_{r})}{(4\pi z)^2} e^{-\gamma}.
\label{eq:LOS}
\end{equation}
Here, $\gamma$ is given by
\begin{equation}
\gamma=\int_{0}^{z} \rho \left\langle \sigma_t \right\rangle \,\, ds,
 \label{eq:gamma_general}
\end{equation}
where the density of the molecules $\rho(s)$ and the total cross-section $\sigma_t(s)$ can be functions of the position along the path from the Tx to the Rx~\cite{ishimaru1978wave}. In specific,  $\sigma_t=\sigma_{abs}+\sigma_{sca}$ is the sum of both the absorption and scattering cross-sections, and $\left\langle\cdot\right\rangle$ represents the average over the size distribution. The absorption cross-section, $\sigma_{abs}$, and the scattering cross-section, $\sigma_{sca}$, have been derived  in~\cite{9838833}.

To take into account the effect of the particle size distribution, we have\begin{equation}
\rho\left\langle \sigma_t\right\rangle=\int_{0}^{\infty}  \sigma_t(x_{d})n(x_d,\mathbf{r}) \,\,dx_{d},
\end{equation} where $n(x_{d},\mathbf{r})$ is the number of RBCs per unit volume located at $\mathbf{r}$ having a range of sizes, in this case, the diameter of the RBC,  between $x_d$ and $x_d + dx_d$. Therefore, the number density distribution as a function of the particle size  follows a log-normal distribution given by~\cite{mclaren1986analysis}
\begin{equation}
n(x_{d},\mathbf{r})=\frac{1}{x_d \sqrt{(2\pi)}\ln (\sigma_g)}e^{-\frac{(\ln(x_d)-\ln(r_g))^2}{2 \ln (\sigma_g)^2}},
\label{eq:number_density}
\end{equation}
where $r_g$ is the mean geometric diameter of the scatterers, and $\sigma_g$ is the corresponding
standard deviation. We note that skewed distributions are particularly common when  values cannot be negative due to the physical aspects of the problem. In~\eqref{eq:LOS}, $\lambda_{g}=\lambda/n'(\omega)$, where $n'(\omega)$ is the real part of the protein refractive index. We compute $n'(\omega)$ using the protein permittivity expression given in~\cite{knab2006hydration}.

\subsection{Path-Loss and Shadowing}
\label{Sec:Sec3B}
   The dynamic nature of intra-body communications results in random shadowing events. These events occur when the transmitted EM waves encounter obstacles, in this case, particles, that obscure the direct LOS signal path. The severity of shadowing depends on factors such as the size and material properties of the obstructing objects, the frequency of the signal, and the distance between the Tx and Rx.
   
   According to the central limit theorem, the attenuation, $X_{\sigma}$, can be modeled as a Gaussian random variable in the log-scale, referred to as log-normal shadowing. Hence, $X_{\sigma}$  has the following distribution  
   \begin{equation}
f(X_{\sigma})=\frac{1}{\sqrt{2\pi\sigma_{X}^2}}e^{-\frac{(X_{\sigma}-\mu_{X})^2}{2\sigma_{X}^2} },
\end{equation}
where $\mu_{X}$ and $\sigma^2_{X}$  are the mean and variance  in dB of the shadowing variable, respectively. In this paper, the parameters defining the pdf of $X_{\sigma}$ are extracted by implementing Monte Carlo simulations~\cite{9838833}. By incorporating the effect of shadowing and knowing the power received from~\eqref{eq:LOS}, the path-loss can be formulated as
\begin{equation}
\left[ L_{LOS}\right]_{\mathbf{dB}}=\left[ \frac{(4\pi z)^2}{\lambda^2_g G_t(\theta_{t})G_r(\theta_{r })} e^{\gamma} \right]_{\mathbf{d}\mathbf{B}}+X_{\sigma}.
\end{equation}

\section{Modeling Protein Dynamics}
\label{Sec:Sec4}
 \subsection{Langevin Equation}
To model protein dynamics, we analyze the forces acting on the system. These forces include impacts from the external force exerted by the nanoantenna, which is subject to channel randomness, as well as forces arising from the Brownian motion of the surrounding particles. By solving the Langevin equation, researchers can gain insights into the behavior and dynamics of proteins, including their diffusion properties, conformational changes, and interactions with their surroundings. Therefore,  we deploy the Langevin stochastic equation, driven by these forces as follows~\cite{mccammon1984protein} \begin{equation}m \frac{d^2 x}{dt^2}+\ddot\beta \frac{dx}{dt}+kx(t)=f_{ex}(t)+f_{\zeta}(t),
 \label{eq:main_0}
\end{equation}
where $x=x(t)$ is the protein coordinate, $m$ is the protein mass, $k$ is the protein spring constant, and $\ddot \beta$ is the damping coefficient. In addition, $f_{ex}(t)$ is the external driving nanoantenna force, which can be represented 
as\begin{equation}
f_{ex}(t)=f_o\cos(\omega_{a} t+\Theta)
\label{eq:random_force},
\end{equation}
while $f_{\zeta}(t)$  is the internal force acting on the protein due to the random motion of the particles suspended in the medium.
\subsubsection{Nanoantenna External Force}
To relate the nanoantenna external force, $f_{ex}(t)$, to the propagation model, we first exploit the relationship between the received power, $P_r$, absorbed by the protein and the impinging nanoantenna electric field strength, $E_o$. This is given as~\cite{cheng1989field}
\begin{equation}
P_r=\frac{1}{2 \mathbf{R}\mathbf{e}\left\{ \eta^{*}_c \right\}}E_{o}^2A_r,
\label{eq:relationship}
\end{equation}
where $\eta^{*}_{c}$ is the intrinsic impedance in a lossy medium,  given by~\cite{cheng1989field}
\begin{equation}
\eta_c=\sqrt{\frac{\ddot\mu}{\varepsilon'_m}}\left( 1-j\frac{\varepsilon''_{m}}{\varepsilon'_m} \right)^{-\frac{1}{2}}.
\end{equation}
In addition, $A_r$ is the receiving cross-section for the THz wave incident on the protein. It is given as 
\begin{equation}
A_r=\frac{\lambda_{g}^2}{4 \pi}G_r(\theta_r).
\label{eq:effective_aperture}
\end{equation}    
We then find the relationship  between the incident electric field and the force stimulating the protein by solving for $E^{2}_o$. This is done by equating~\eqref{eq:LOS} to~\eqref{eq:relationship}, which yields 
\begin{equation}
E^2_o=\frac{2P_tG_t(\theta_t)G_r(\theta_r) \mathbf{R}\mathbf{e}\left\{ \eta^{*}_c \right\}}{L_{LOS} \, A_{r}}.
\label{electric_field}
\end{equation}
Finally, we substitute~\eqref{electric_field} in~\eqref{eq:force_eq} to obtain
\begin{equation}
f^2_{o}=\left( \frac{-4\pi  r^{3} \varepsilon_m \mathbf{Re}\left\{ \mathcal{K}\eta^{*}_c  \right\}\kappa P_{t}G_t(\theta_t) G_r(\theta_r)  e^{-2\kappa z}  }{L_{LOS} A_r} \right)^2.
\label{eq:force_ff}
\end{equation}
We note here that the randomness in the nanoantenna force stems from the path-loss, $L_{LOS}$, experienced by the channel due to the large number of molecules.

\begin{figure}[h!]
\centering
\includegraphics[width=0.4\textwidth]{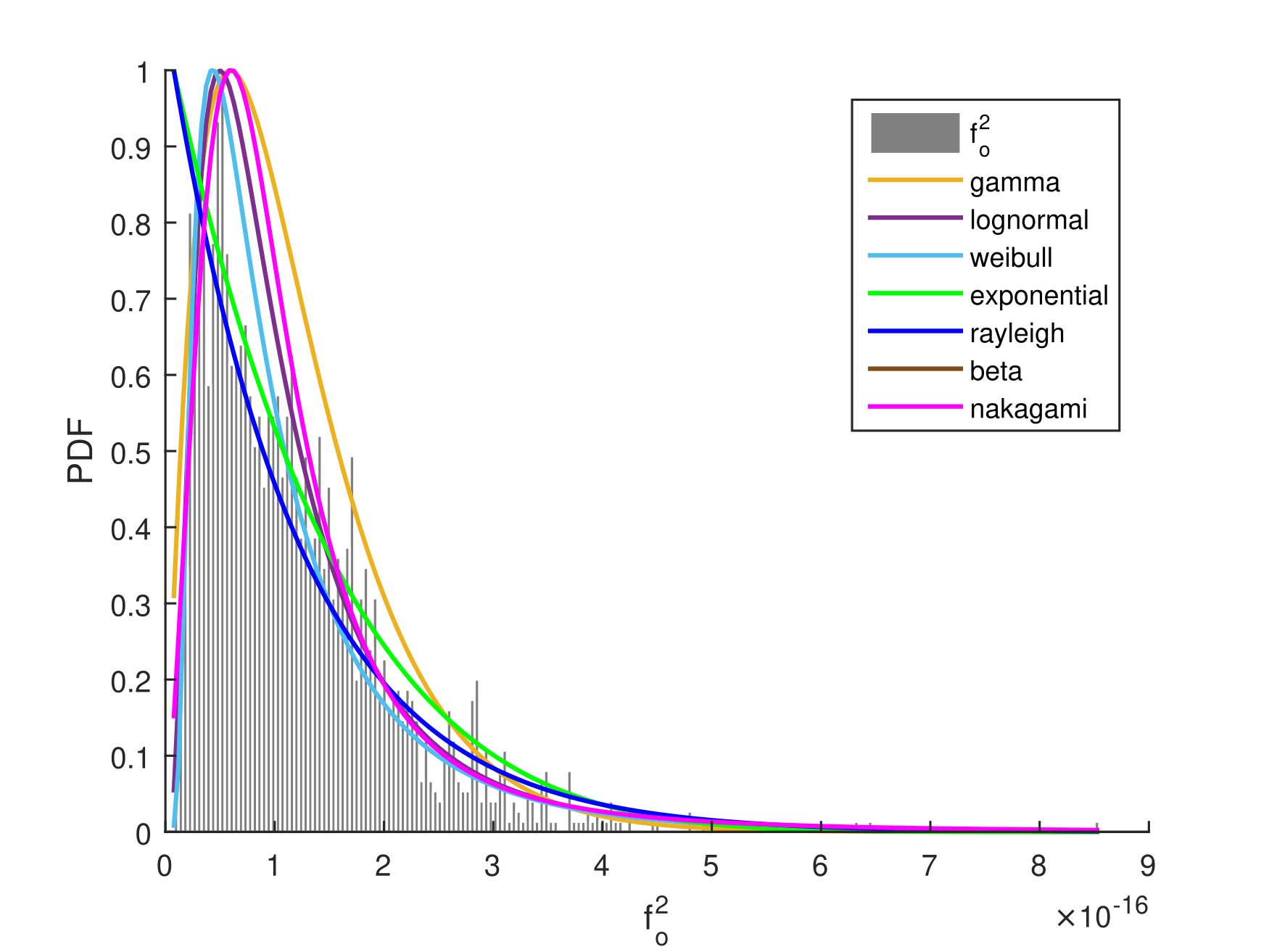}
\footnotesize
\caption{Fitting $f^2_o$ values obtained from the received EM field strength in COMSOL, where $\alpha=2.5\times10^{-17}$ and $\beta=1.9$.}
\label{fig:fitting}
\end{figure}

Using COMSOL Multiphysics$\textsuperscript{\textregistered}$ software, we built a physical model that mimics our intra-body scenario and fitted $f^2_o$ as demonstrated in Fig.~\ref{fig:fitting}. Among the candidate distributions, the gamma distribution was the best fit, as indicated by the log-likelihood value provided in Table~\ref{table:distribution}. The log likelihood criterion  is a measure of goodness of fit for any model, in which the higher the log-likelihood value, the better the model.

\begin{table}[h!]
\caption{Comparison of distributions for the best-fit of  $f^2_o$}\begin{center}
\centering
\small
\label{table:distribution}
\begin{tabular}{|p{45pt}|p{40pt}|p{65pt}|}\hline
\begin{footnotesize}\textbf{Distribution} \end{footnotesize}&\begin{footnotesize} \textbf{Best-Fit} \end{footnotesize}&\textbf{\begin{footnotesize}Log likelihood \textbf{} \end{footnotesize}} \\\hline
 \begin{footnotesize}Gamma  \end{footnotesize}&\begin{footnotesize} 1 \end{footnotesize} &\begin{footnotesize} 7.167e+04\end{footnotesize}\\\hline
\begin{footnotesize}Lognormal \end{footnotesize}& \begin{footnotesize}  2 \end{footnotesize}&\begin{footnotesize} 7.163e+04 \end{footnotesize}\\\hline
\begin{footnotesize}Weibull \end{footnotesize}& \begin{footnotesize}  3 \end{footnotesize}&\begin{footnotesize} 7.161e+04\end{footnotesize}\\\hline
 \begin{footnotesize}Nakagami \end{footnotesize}& \begin{footnotesize} 4  \end{footnotesize}&\begin{footnotesize} 7.155e+04 \end{footnotesize}\\\hline
 \begin{footnotesize}Exponential \end{footnotesize}& \begin{footnotesize}  5 \end{footnotesize}&\begin{footnotesize} 7.143e+04 \end{footnotesize}\\\hline
  
  \begin{footnotesize}Rayleigh \end{footnotesize}& \begin{footnotesize}  6 \end{footnotesize}&\begin{footnotesize}  7.135e+04 \end{footnotesize}\\\hline
\end{tabular}
\end{center}
\end{table}

  Thereby, $f^2_o$ follows a gamma distribution given by 
\begin{equation}
g(f_o^{2}; \alpha,\beta)=\frac{f_o^{2(\alpha-1)}e^{-\beta f^2_{o}} \beta^{\alpha}}{ \Gamma(\alpha)}.
\end{equation}The gamma distribution is a two-parameter family of continuous probability distributions, comprising a shape parameter $\alpha$ and a rate parameter $\beta$. Changing $\alpha$ alters the shape of the probability density function, while adjusting $\beta$ either stretches or compresses the range of the gamma distribution. It's worth noting that varying the density, $\rho$, and size, $x_d$, of the RBCs results in different values for $\alpha$ and $\beta$.

When $\alpha=1$, the resulting gamma distribution is equivalent to the exponential distribution. Interestingly, single exponential decays serve as a natural starting point for analyzing protein relaxation phenomena, as they are phenomenological descriptors of simple dashpots—dampers that exhibit frictional resistance to motion. In addition, the distribution of protein abundances in a cell is well described in the
literature by a gamma distribution~\cite{cai2006stochastic}.

Moreover, $\Theta$ in~\eqref{eq:random_force} is a random variable uniformly distributed between~$\left[-\pi,\pi\right]$, and $\omega_a$ is the nanoantenna frequency. Since the amplitude and phase in~\eqref{eq:random_force} are independent, we can prove that $f_{ex}(t)$ is a wide-sense stationary process, by computing its mean and autocorrelation function. For a gamma distribution, the mean and variance are given by  $E[f^2_o]=\frac{\alpha}{\beta}$ and $Var[f^2_o]=\frac{\alpha}{\beta^2}$, respectively. The auto-correlation function, for time-lag $\tau$,  yields
\begin{equation}
R_{f_{ex}}(\tau)=\frac{\alpha}{2 \beta}\cos(\omega_{a}\tau). 
\end{equation}

\subsubsection{Force due to Brownian Motion}
 In~\eqref{eq:main_0}, the internal stochastic force $f_{\zeta}(t)$ is governed by a white-noise fluctuation-dissipation relation as follows~\cite{balakrishnan2008elements}
\begin{equation}
f_{\zeta}(t)={\sqrt{\Gamma}}\zeta(t).
\label{eq:s_f}
\end{equation}
Here, $\zeta(t)$ is a white, Gaussian, random process with moments~\cite{balakrishnan2008elements}
\begin{equation} 
\begin{split}
\langle \zeta(t_1)\rangle&=0 \\
\langle \zeta(t_1)\zeta(t_2)\rangle&=\delta(t_1-t_2).
\label{eq:moments}
\end{split}
\end{equation}
In~\eqref{eq:s_f}, $\Gamma= 2\bar\gamma mk_bT$ ($k_b$ is Boltzmann's constant and $T$ is the temperature in Kelvin) denotes the strength of the noise $\zeta(t)$. It fixes the amplitude of the fluctuation in the random force in terms of both the temperature  and the dissipation coefficient~\cite{balakrishnan2008elements}.

The Green's function is a powerful mathematical tool to solve inhomogeneous differential equations.  For an arbitrary forcing term, the  solution to the  equation is formulated  by integrating the Green's function against the forcing term. To solve~\eqref{eq:main_0}, we first re-write it as
 \begin{equation} \frac{d^2 x}{dt^2}+ \bar\gamma\frac{dx}{dt}+\omega_o^2x(t)=\frac{1}{m}\left[ f_{ex}(t)+f_{\zeta}(t) \right], 
\label{eq:main_1}
\end{equation}
where  $\bar \gamma={\ddot\beta}/{m}$ and $\omega_o^2=k/m$~\cite{carpinteri2017terahertz}. Here, $\omega_o$ is  the natural frequency of the protein, and $\bar\gamma$ is the damping constant that governs the magnitude of the protein vibrational resonances.

In the context of modeling proteins, the interatomic forces are often approximated using harmonic potentials. This approximation is based on the assumption that the potential energy between atoms within the protein can be described as a harmonic oscillator potential. In fact, in~\cite{tirion1996large}, the author showed that the vibrational modes of protein molecules are not significantly modified when interactions are replaced by Hookean springs, for all atom pairs whose distance is smaller than a given cut-off value.

The frequency-domain solution of the Green's function corresponding to~\eqref{eq:main_1} is found as~\cite{byron2012mathematics} 
\begin{equation}
G(\omega)=\frac{1}{m\left[ -\omega^2- j\bar\gamma \omega+\omega^2_o \right]},
\label{eq:gfd}
\end{equation}
where $G(\omega)$ represents the transfer function of the protein dynamics. The solution of~\eqref{eq:main_1} can be decomposed into two parts, one related to the  nanoantenna external force $f_{ex}(t)$ and the other is related to the stochastic force $f_{\zeta}(t)$, resulting in $x(t)=~x_{ex}(t)+x_{\zeta}(t)$. 

In our model, we consider $\omega_o>\frac{\bar\gamma}{2}$. Physically, this condition satisfies protein collective vibrations, where protein
modes in the THz frequency range have been shown to be underdamped even in aqueous solutions~\cite{turton2014terahertz,acbas2014optical}. 


\subsection{Energy of Driven Protein}

Proteins are dynamic entities capable of adopting various conformations to perform their biological functions. These conformational changes are often triggered by the exchange of energy between the protein and its environment. The requirement for energy transfer to induce a conformational change highlights the importance of connecting THz band signals with protein molecules to initiate resonance. When an external harmonic excitation has a frequency that matches one of the natural frequencies of the system, resonance occurs, leading to an increase in the vibrational amplitude of the structure~\cite{carpinteri2017terahertz}.

In protein folding, the energy landscape and the resulting folding pathway play a crucial role in determining the probability of a protein adopting its native conformation. To assess the impact of THz signaling on protein folding probability, we must first calculate the total energy absorbed by the protein when subjected to both the nanoantenna force and the force resulting from the surrounding Brownian motion. To do so, we sum the kinetic and potential energy contents~\cite{elayan2021enabling} 
\begin{equation}
\left\langle E_{tot} \right\rangle_{ss}=\frac{1}{2}m \left\langle v^2(t) \right\rangle_{ss} +\frac{1}{2}k \left\langle x^2(t) \right\rangle_{ss}.
\label{eq:driving_energy}
\end{equation}We denote by $\left\langle\cdot \right\rangle_{ss}$ the  statistical  average value in steady-state, where all transient effects die out. 

To evaluate the potential energy,  we find the average steady-state squared displacement of the protein, $\left\langle x^2(t) \right\rangle_{ss}$, which is decomposed into
\begin{equation}
\left\langle x^2(t) \right\rangle_{ss}=\left\langle x_{ex}^2(t)\right\rangle_{ss}+\left\langle x_{\zeta}^2(t)\right\rangle_{ss}.
\label{eq:x_ss}
\end{equation}
$\left\langle x^2(t) \right\rangle_{ss}$  is a measure of the spatial extent of the protein motion. As detailed in~\cite{elayan2022selectivity}, $\left\langle x_{\zeta}^2(t)\right\rangle_{ss}$ yields $\frac{k_bT}{k}$. In terms of $\left\langle x_{ex}^2(t)\right\rangle_{ss}$, we start by examining  $f_{ex}(t)$ given in~\eqref{eq:random_force}. Since the latter is a wide-sense stationary random process with autocorrelation
function $R_{f_{ex}}(\tau)$, its power spectral density (PSD) $S_{{ex}}(\omega)$ is introduced as the Fourier transform of $R_{f_{ex}}(\tau)$. This relationship is given by
\begin{equation}
\begin{split}
S_{{ex}}(\omega)&=\int_{-\infty}^{\infty}R_{f_{ex}}(\tau)e^{-j \omega \tau} d \tau \\
&=\frac{\pi\alpha}{2 \beta}\left[ \delta (\omega-\omega_a)+\delta(\omega+\omega_a) \right].
\end{split}
\label{eq:extent1}
\end{equation} 
Using the PSD of the  driving force and the frequency-domain solution of  Green's function in~\eqref{eq:gfd}, the output PSD is given by 
\begin{equation}
S_{x}(\omega)=|G(\omega)|^{2} \cdot S_{{ex}}(\omega).
\label{eq:PSDex}
\end{equation}
Finally, from~\eqref{eq:PSDex}, the average squared displacement as a result of the nanoantenna force, $\left\langle x^2_{ex}(t) \right\rangle_{ss}$, is   found as \begin{equation}
\begin{split}
\left\langle x^2_{ex}(t) \right\rangle_{ss}&=\frac{1}{2\pi}\int_{-\infty}^{\infty} S_{x}(\omega) d\omega\\
&=\frac{1}{2\pi}\frac{\pi \alpha}{2 \beta }\int_{-\infty}^{\infty}\frac{(\delta (\omega-\omega_a)+\delta(\omega+\omega_a))}{m^{2}[(\omega_{o}^2-\omega^2)^{2}+(\omega\bar\gamma)^2]}
d\omega \\
&=\frac{\alpha}{2 \beta m^{2}[(\omega_{o}^2-\omega_{a}^2)^{2}+(\omega_{a}\bar\gamma)^2]}.
\end{split}
\label{eq:disp_0}
\end{equation}

To calculate the kinetic energy, we need to compute the  average steady-state squared velocity of the protein, $\left\langle v^2(t) \right\rangle_{ss}$, which is decomposed into 
\begin{equation}
\left\langle v^2(t) \right\rangle_{ss}=\left\langle v_{ex}^2(t) \right\rangle_{ss}+\left\langle v_{\zeta}^2(t)\right\rangle_{ss}.
\label{eq:v_ss}
\end{equation} 
From the derivation in~\cite{elayan2022selectivity}, we know that $\left\langle v_{\zeta}^2(t)\right\rangle_{ss}$ yields $\frac{k_bT}{m}$. Further, the steady-state average squared velocity resulting from  the nanoantenna force can be found from the Fourier transform of the derivative of the displacement. From~\eqref{eq:disp_0}, we have 
\begin{equation}
\begin{split}
\left\langle v^2_{ex}(t)\right\rangle_{ss}&=\frac{1}{2\pi}\int_{-\infty}^{\infty}S_{x}(\omega)\omega^2 d\omega\\
&=\frac{1}{2\pi}\frac{\alpha\pi}{2 \beta}\int_{-\infty}^{\infty}\frac{ \omega^2(\delta (\omega-\omega_a)+\delta(\omega+\omega_a))}{m^2[\left(\omega_{o}^2-\omega^2)^{2}+(\omega\bar\gamma\right)^2]}
d\omega \\
&=\frac{\alpha \omega^2_a}{2\beta m^{2}[(\omega_{o}^2-\omega_{a}^2)^{2}+(\omega_{a}\bar\gamma)^2]}.
\end{split}
\label{eq:part2b}
\end{equation}

Finally, we find the total steady-state energy of the driven protein motion  by substituting~\eqref{eq:x_ss} and~\eqref{eq:v_ss} in~\eqref{eq:driving_energy} to yield
\begin{equation}
\begin{split}
\left\langle E_{tot}\right\rangle _{ss}
=&\underbrace{\frac{\alpha}{4\beta m[(\omega_{o}^2-\omega_{a}^2)^{2}+(\omega_{a}\bar\gamma)^2]}( \omega_{a}^{2}+\omega^2_o)}_{\text{THz  Contribution}} +\underbrace{k_{b}T}_{\text{Noise Effect}}.
\end{split}
\label{eq:average_energy1}
\end{equation}
If we compare $\left\langle E_{\text{tot}}\right\rangle _{\text{ss}}$ in~\eqref{eq:average_energy1} to our previous work in~\cite{elayan2022selectivity}, we can observe the influence of the gamma distribution parameters, specifically $\alpha$ and $\beta$, on the energy expression. $\alpha$ represents the amplitude of the squared force, while $\beta$ governs the extent of path-loss. This clearly underscores the influence of the channel on the incident nanoantenna force, with the values of $\alpha$ and $\beta$ determining the amount of energy absorbed by the protein particle.

\subsection{ Boltzmann Distribution}
Energy functions are linked to molecular conformation through the Boltzmann distribution. This distribution provides the probability of a system being in a specific state based on the energy of that state and the system's temperature~\cite{finkelstein1995protein}. In the context of protein folding, the Boltzmann distribution aids in understanding the equilibrium populations of different folding states. At a given temperature, protein states with lower energies are more likely to be populated, while states with higher energies have lower probabilities. The state with the lowest energy corresponds to the protein's native conformation, while higher energy states represent various unfolded or partially folded conformations.

In this work, we model the protein as having two states, unfolded and folded. Through the  folding structure, the protein acquires a conformation that is biologically functional.  The rate of protein folding is given by~\cite{sachs1991stochastic}
\begin{equation}
r_{f}=r_0\exp \left(\frac{-E_f}{k_bT}\right),
\label{eq:fr0}
\end{equation}
where $E_f$ denotes the  free-energy associated with the folded protein state, and $r_0$ is a scale factor which preserves detailed balance. In our case, we amend~\eqref{eq:fr0} to  incorporate~\eqref{eq:average_energy1}  as follows\begin{equation}
r_{f}=r_0\exp \left(\frac{-E_f+\langle E_{tot}\rangle_{ss}}{k_bT}\right).
\label{eq:fr}
\end{equation}
The unfolding transition is a relaxation process, which returns
the protein to the unfolded state. This process is considered independent of the imposed signal~\cite{anfinsen1973principles} and is given by
\begin{equation}
r_{u}=r_0\exp \left(\frac{-E_u}{k_bT}\right),
\label{eq:br}
\end{equation} where $E_u$ denotes the  free-energy associated with the unfolded protein state. Consequently, the rate of change in the protein folding probability is given by
\begin{equation}
\frac{d}{dt}p_{f}(t)=-r_{u}p_{f}(t) +r_{f} (1-p_{f}(t)).
\label{eq:master_eq}
\end{equation}
Here, $p_{f}(t)$ is the probability that
the protein is in the folded  state at time $t$ (the corresponding probability of the unfolded state is given by $p_{u}(t)=1-p_{f}(t)$). From (\ref{eq:master_eq}), the steady-state solution can be found as
\begin{equation}
p_{F}=\frac{r_f}{r_f+r_u}.
\label{eq:pf}
\end{equation}
Substituting \eqref{eq:fr} and \eqref{eq:br} in \eqref{eq:pf} yields 
\begin{equation}
p_{F}=\frac{1}{1+\exp\left(\frac{\Delta E-\langle E_{tot}\rangle_{ss}}{k_{b}T}\right)},
\label{eq:pf_ff}
\end{equation}
where $\Delta E=E_f-E_u$ is the protein free-energy.  

 The state of a single protein  can be regarded as a Bernoulli random variable with a probability of success, $p_{F}$. When considering a system composed of $n$ proteins, the random number of folded proteins, $n_F$, follows a binomial distribution. Due to the large diversity of proteins seen in an intra-body environment, the binomial distribution can be approximated by a normal distribution with mean, $\mu=np_{F}$, and variance, $\sigma^2=np_{F}(1-p_{F})$. Finally, 
we define protein population as the number of copies of a protein molecule in a cell. Hence, the number of folded proteins of each  protein population, $i$, can be expressed as a normal distribution as follows 
\begin{equation}
n_{F,i}\sim\mathcal{N}\left(np_{F,i}, np_{F,i}(1-p_{F,i})\right).
\end{equation}

\section{The Impact of Fading on THz-Induced Protein Interactions}
\label{Sec:Sec5}
Fading refers to the attenuation of the THz nanoantenna signal caused by the presence of RBCs. Furthermore, due to the mobility of RBCs, fading varies over time. This variation has implications for controlling the dynamics of the desired protein population and places certain requirements on the system parameters. To understand the impact of the channel on the receiver, we need to determine how quickly the communication channel changes relative to the protein folding time. Therefore, we use the coherence time, denoted as $T_c$, as a metric to characterize the nature of the channel in the time domain~\cite{tse2005fundamentals}.
$T_c$ is inversely proportional to the maximum Doppler spread, $f_m$. We have $T_c\approx\frac{1}{f_m}$, where $f_m$ is computed using~\cite{tse2005fundamentals}
\begin{equation}f_m=\frac{v}{c}f_c.
\end{equation} Here, $f_c$ is the  center frequency of the nanoantenna (in our case it is the resonant frequency of the protein of interest), $v$ is the velocity of RBCs, and $c$ is the speed of light.

 \textit{Fast fading} occurs when the channel impulse response
changes rapidly within the symbol duration $T_s$.  In this case,  high Doppler spread is observed,  where $T_c\ll T_s$~\cite{tse2005fundamentals}, and $T_s$ represents the protein folding time. Since fast fading  is characterized by the rapid fluctuations of the signal over small distances, it is considered an ergodic process that allows the use of average metrics for its assessment. Therefore, to compute the  probability of the protein in a folded state,  we use the form of the probability $p_F$  given in~\eqref{eq:pf_ff}. Basically,  $\langle E_{tot}\rangle_{ss}$ in~\eqref{eq:pf_ff} represents the averaged energy, which  captures the incident wave randomness that  is changing very fast.
 
On the other hand, \textit{slow fading} arises when the channel impulse response is considered roughly constant over the period of use~\cite{tse2005fundamentals}. In slow fading, low  Doppler spread is observed, where $T_c \gg T_s$. Slow fading is a long-term fading effect that changes the mean value of the received signal. To compute the  probability of the protein in a folded state under such conditions, we  take the expectation of $p_F$ as 
\begin{equation}
\begin{split}
E[p_F]&=\int_{0}^{\infty} \frac{1}{1+\exp(\frac{\Delta E-E_{i}}{k_bT})}g( E_i; \alpha,\beta)\,\, d E_{i},
\end{split}
\label{eq:incident_energy}
\end{equation}
where the instantaneous energy $E_i$ follows a  gamma distribution, $g(E_i; \alpha,\beta)$, stemming from the fact that $E_i\propto f^2_o$. 
 
 In~\eqref{eq:incident_energy}, $E_i=f_o \times \Delta x$ results from  the interaction between the nanoantenna force, $f_{o}$, and the protein molecule, where $\Delta x$  corresponds to a conformational change in the protein structure. From the relationship between $f_o$, $\Delta x$ and the protein stiffness $k$, we can write $\Delta x=f_o/k$. Finally, from the vibrational frequency of the protein structure, we have  $k=m\omega^2_o$~\cite{carpinteri2017terahertz}. The instantaneous energy therefore  yields \begin{equation} 
E_i=\frac{f^2_o}{m \omega^2_o},
\end{equation}
where $f^2_o$ is given in~\eqref{eq:force_ff}.

\subsection{ Selectivity in Fading Scenarios }

\textit{Selectivity} refers to the nanoantenna's ability to induce a conformational change in the desired protein population without affecting the conformation of other untargeted proteins in the system. It is a metric that assesses the controllability of THz signals over protein networks. In our previous work~\cite{elayan2022selectivity}, we proposed the selectivity metric as
\begin{equation}
S({\mu_d}, \mu_{ud})=\frac{\mu_d-\mu_{ud}}{\max (\mu_d,\mu_{ud})},
\label{eq:metric2}
\end{equation}
where $\mu_d=n_{d}p_{F,d}$ and $\mu_{ud}=n_{un}p_{F,ud}$ are the means of the desired and undesired protein populations, respectively.  The developed selectivity metric has several properties which makes it a powerful tool for evaluating protein interactions in the system as discussed in~\cite{elayan2022selectivity}.  Most importantly, as the metric relies on the stationary probability difference, it encompasses both the system's mechanical parameters and the energy stored in the THz-stimulated protein.

Nevertheless, the presence of fading has an impact on the system's selectivity by potentially causing the reception of a weak signal. This weak signal may lead to a temporary disruption in the communication link between the nanoantenna and the target protein due to a significant decrease in power. This, in turn, affects the system's ability to target the desired protein population. Although fading in an intra-body scenario is characterized by its unpredictability and stochastic nature, our metric can still be optimized to ensure that the system achieves the highest level of selectivity possible. Understanding the interactions between proteins and components within the bloodstream, and how these interactions influence disease progression, can aid in identifying potential therapeutic targets. By concentrating on proteins involved in specific pathways, researchers can develop therapies that selectively modulate processes associated with the disease.

In our model,  the nanoantenna Tx power, $P_t$, and  frequency, $\omega_a$ are the only system parameters that can be controlled. We, therefore, formulate a joint optimization problem to maximize the selectivity  with respect to $P_t$ and  $\omega_a$. In a fast-fading scenario, the optimization problem yields
\begin{equation}
\begin{aligned}
& \underset{P_{t},\text{$\omega_a$}}{\text{max.} \,\,}
& & \text{$S(\mu_{d},\mu_{ud})$} \\
& \text{subject to}
& & 0\leq p_{F,d}\leq  \ 1 \\
& & & 0\leq p_{F,ud}\leq \ 1. \\
\end{aligned}
 \label{eq:sys_const1}
\end{equation}
The optimization problem is first solved by expressing $p_F$ in terms of $P_{t}$ and $\omega_a$. Then, using  the fixed point iteration method~\cite{cegielski2012iterative}, we  numerically solve the joint optimization problem in~\eqref{eq:sys_const1}, where the optimal values  that  maximize the selectivity are retrieved.

\subsection{Probability of Selectivity Outage}
  In the  case of slow fading,  the selectivity is given as 

\begin{equation}
\begin{split}
S(E[\mu_{d}],E[\mu_{ud}])&=\frac{E[\mu_{d}]-E[\mu_{ud}]}{\max (E[\mu_{d}],E[\mu_{ud}])},
\end{split}
\label{eq:metric3}
\end{equation}where $E[\mu]=nE[p_F]$ and $E[p_F]$ is solved using~\eqref{eq:incident_energy}.  The optimization problem is formulated  as 
\begin{equation}
\begin{aligned}
& \underset{P_{t},\text{$\omega_a$}}{\text{max.} \,\,}
& & \text{$S(E[\mu_{d}],E[\mu_{ud}])$} \\
& \text{subject to}
& & 0\leq  p_{F,d}\leq  \ 1 \\
& & & 0\leq p_{F,ud}\leq \ 1. \\
\end{aligned}
\label{eq:metric3_optimize}
\end{equation}Similar to the previous case,~\eqref{eq:metric3_optimize} is numerically solved using fixed point iteration and the optimal $P_t$ and $\omega_a$ are found.

It's important to note that in conditions of slow fading, the wireless channel is categorized as non-ergodic, meaning that the capacity of the channel becomes a random variable~\cite{tse2005fundamentals}. In such scenarios, a measure of the channel's quality is the \textit{ outage probability}, which represents the likelihood of the information rate falling below the required threshold rate. To maintain acceptable communication performance, a minimum signal level is necessary, ensuring that the received signal maintains sufficient strength during ``non-fade intervals". However, if the signal level drops below this threshold, the system will experience insufficient signal strength, resulting in what is commonly referred to as an ``outage."

The concept of ergodicity is often employed in the context of protein dynamics to describe the statistical properties of conformational changes over time. However, certain protein systems exhibit non-ergodic behavior, indicating that their conformational dynamics cannot be easily sampled or described by a single trajectory or ensemble of conformations. Non-ergodicity and slow fading are related in the sense that slow fading can lead to non-ergodic behavior in wireless communication channels. To study slow fading in an intra-body scenario, we define \textit{selectivity outage} as a situation where the selectivity falls below a certain threshold, i.e., $S(E[\mu_{d}],E[\mu_{ud}])<\gamma_o$. Selectivity can drop below the defined threshold when either the desired protein population receives a minimal amount of power, insufficient for folding the targeted proteins, or when the system receives a substantial amount of power, causing all proteins to fold. In both cases, we lose the ability to target a specific protein.

In this context, we introduce the \textit{probability of selectivity outage} as the probability of attaining a selectivity below the defined threshold, i.e., $\mathrm{Pr}(S(E[\mu_{d}],E[\mu_{ud}])<\gamma_o)$. A trade-off must  exist between the selectivity threshold and the outage probability since the  higher the threshold, the higher the probability of selectivity outage. When an outage occurs in the system, the nanoantenna must re-transmit the signal. This not only  results in unconstrained delays but also sacrifices the system's energy by consuming more power. Different medical applications often require varying degrees of selectivity depending on the specific goals and requirements of the application. Therefore, the application must dictate whether or not the system can sustain being in an outage.   

The probability of selectivity outage serves as a metric for assessing the quality of nanoantenna transmission in an intra-body network under slow fading conditions. By setting a threshold value, the best performance achievable by the nanoantenna is to maintain a selectivity level equal to that threshold. Reliable controllability is achieved when this threshold is met; otherwise, an outage occurs.

 \section{ Results}
\label{Sec:Sec6}

\begin{table}
\caption{Simulation Parameters}\begin{center}
\centering
\small
\label{table:parameters}
\begin{tabular}{|p{122pt}|p{64pt}|p{24pt}|}\hline
\begin{footnotesize}\textbf{Parameters} \end{footnotesize}&\begin{footnotesize} \textbf{Value} \end{footnotesize}& \begin{footnotesize} \textbf{Ref.} \end{footnotesize}\\\hline
 \begin{footnotesize}Rhodopsin resonant  frequency   \end{footnotesize}&\begin{footnotesize} 1.36 THz\end{footnotesize} &\begin{footnotesize}\cite{balu2008terahertz}\end{footnotesize} \\\hline
 \begin{footnotesize} Rhodopsin stiffness\end{footnotesize}& \begin{footnotesize}3 N/m\end{footnotesize}&\begin{footnotesize}\cite{sapra2008mechanical}\end{footnotesize} \\\hline
\begin{footnotesize}Rhodopsin mass \end{footnotesize}& \begin{footnotesize}$1.62\times10^{-24}$ kg\end{footnotesize}&\begin{footnotesize}-\end{footnotesize}\\\hline
\begin{footnotesize}Bacteriorhodopsin resonant frequency \end{footnotesize}& \begin{footnotesize}1.13 THz\end{footnotesize}&\begin{footnotesize}\cite{balu2008terahertz}\end{footnotesize}\\\hline
\begin{footnotesize}Bacteriorhodopsin stiffness \end{footnotesize}& \begin{footnotesize}1.9 N/m\end{footnotesize}&\begin{footnotesize}\cite{sapra2008mechanical}\end{footnotesize}\\\hline
\begin{footnotesize}Bacteriorhodopsin mass \end{footnotesize}& \begin{footnotesize}$1.48\times10^{-24}$ kg\end{footnotesize}&\begin{footnotesize}-\end{footnotesize}\\\hline
\begin{footnotesize}Damping constant ($\bar \gamma)$\end{footnotesize}& \begin{footnotesize}0.3 THz\end{footnotesize}&\begin{footnotesize}\cite{turton2014terahertz}\end{footnotesize}\\\hline
\begin{footnotesize}Free energy ($\Delta E$)\end{footnotesize}& \begin{footnotesize}6 kbT\end{footnotesize}&\begin{footnotesize}\cite{benham2009mathematics}\end{footnotesize}\\\hline
\begin{footnotesize}Tx Power ($P_t$)\end{footnotesize}& \begin{footnotesize}35 $\mu$W\end{footnotesize}&\begin{footnotesize}-\end{footnotesize}\\\hline
\begin{footnotesize}Tx Gain ($G_t$)\end{footnotesize}& \begin{footnotesize}1.5 (1.76 dBi)\end{footnotesize}&\begin{footnotesize}-\end{footnotesize}\\\hline
\begin{footnotesize}Protein radius ($r$) \end{footnotesize}& \begin{footnotesize}5 nm\end{footnotesize}&\begin{footnotesize}-\end{footnotesize}\\\hline
\begin{footnotesize}Particle density ($\rho$) \end{footnotesize}& \begin{footnotesize}1000 particles/mm$^3$\end{footnotesize}&\begin{footnotesize}-\end{footnotesize}\\\hline
\end{tabular}
\end{center}
\end{table}

\subsection{COMSOL Model} 
We constructed a physical system that replicates our intra-body scenario using COMSOL Multiphysics$\textsuperscript{\textregistered}$. For our analysis, we utilized the simulation parameters provided in Table~\ref{table:parameters}. Specifically, our model focuses on the transmission of an EM wave from a nanoantenna to the protein population of interest within a segment of a blood vessel. In our representation, we employed the \textit{Electromagnetic Waves, Frequency domain interface} to depict the nanoantenna as a short dipole, which is a realistic assumption for short-range communication between nanomachines~\cite{johari2017nanoscale}. We configured the nanoantenna's frequency to match the vibrational frequency of the target protein, rhodopsin (i.e., $1.36$~THz). The protein's radius was set at $5$~nm, and we considered a population of 1000 proteins arranged in a cylindrical shape. Furthermore, we incorporated random particles that mimic RBCs into the field. These particles have a radius of $4~\mu$m and dielectric properties as described in~\cite{reid2013terahertz}.

\begin{figure}[h!]
\centering
\includegraphics[width=0.36\textwidth]{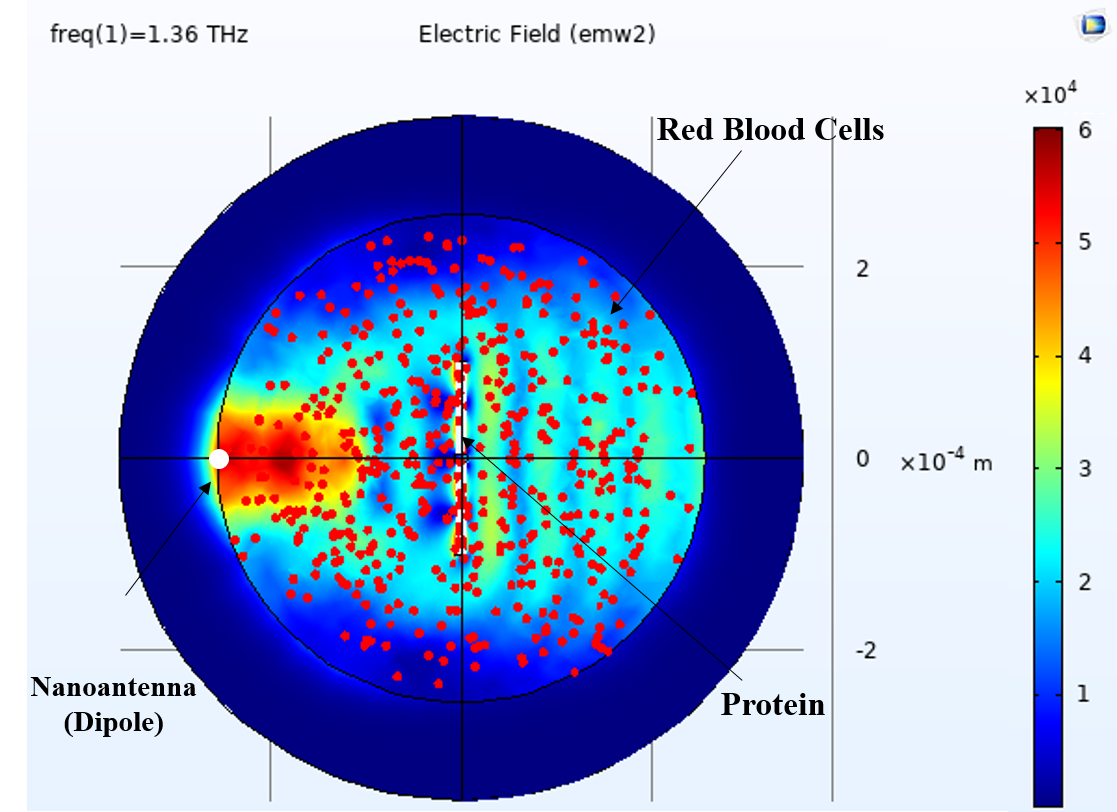}
\footnotesize
\caption{Electric field pattern in V/m for propagation at 1.36~THz. }
\label{fig:COMSOL_mod}
\end{figure}

Fig.~\ref{fig:COMSOL_mod} illustrates the electric field strength received by the targeted protein population at a distance  of $z=500$~$\mu$m, ensuring that we are operating in the far-field.   From the obtained electric field values, we  calculated the force impinging on the protein particles using~\eqref{eq:force_ff}. Then, by sampling the force realizations every $2$~ms,  we found the distribution of the force that  best fits the data based on  a maximum likelihood estimation approach. 

The objective of this work is to demonstrate the influence of a channel composed of a large number of moving particles on protein folding probability. This is accomplished by determining the values of both the steady-state energy, $\langle E_{tot}\rangle_{ss}$, and the instantaneous energy, $E_i$, required to compute $p_F$ in~\eqref{eq:pf_ff} and $E[p_F]$ in~\eqref{eq:incident_energy}, respectively. To create a dynamic environment, we utilized the \textit{Particle Tracing for Fluid Flow} module in COMSOL. We varied the velocity of RBCs to establish a time-varying channel within the range of $0$-$150$ milliseconds (ms). Notably, according to the Protein Folding Kinetics Database, the folding timescales of most proteins are on the order of milliseconds, with a median of $5$ ms for two-state proteins~\cite{manavalan2019pfdb}. Therefore, we set the protein folding time in our system to $T_s=5$ ms. Then, we calculated $T_c$ as summarized in Table~\ref{table:fading_scenarios}. It's worth noting that the speed of RBCs reaches approximately $2$-$3$ mm/s in brain capillaries, $10$ mm/s in brain arterioles, and exceeds $20$-$30$ mm/s in larger pial vessels. Standard scanning systems typically provide scanning velocities ranging from $5$ to $20$ mm/s~\cite{chaigneau2019unbiased}.
 
\begin{table}
\caption{Fading scenarios in an intra-body environment.}\begin{center}
\centering
\small
\label{table:fading_scenarios}
\begin{tabular}{|p{70pt}|p{60pt}|p{88pt}|}\hline
\begin{footnotesize}\textbf{Scenario} \end{footnotesize}&\begin{footnotesize} \textbf{Velocity of RBCs} \end{footnotesize}& \begin{footnotesize} \textbf{Channel Coherence Time}  \end{footnotesize} \\\hline
 \begin{footnotesize}Fast Fading  \end{footnotesize}&\begin{footnotesize} 200  mm/s \end{footnotesize} & \begin{footnotesize}$T_c=1$~ms \end{footnotesize} \\\hline
 \begin{footnotesize}Slow Fading \end{footnotesize}& \begin{footnotesize}5 mm/s   \end{footnotesize}& \begin{footnotesize}$T_c=44$ ms \end{footnotesize} \\\hline
\end{tabular}
\end{center}
\end{table}

\begin{figure}[h!]
\centering
\includegraphics[width=0.4\textwidth]{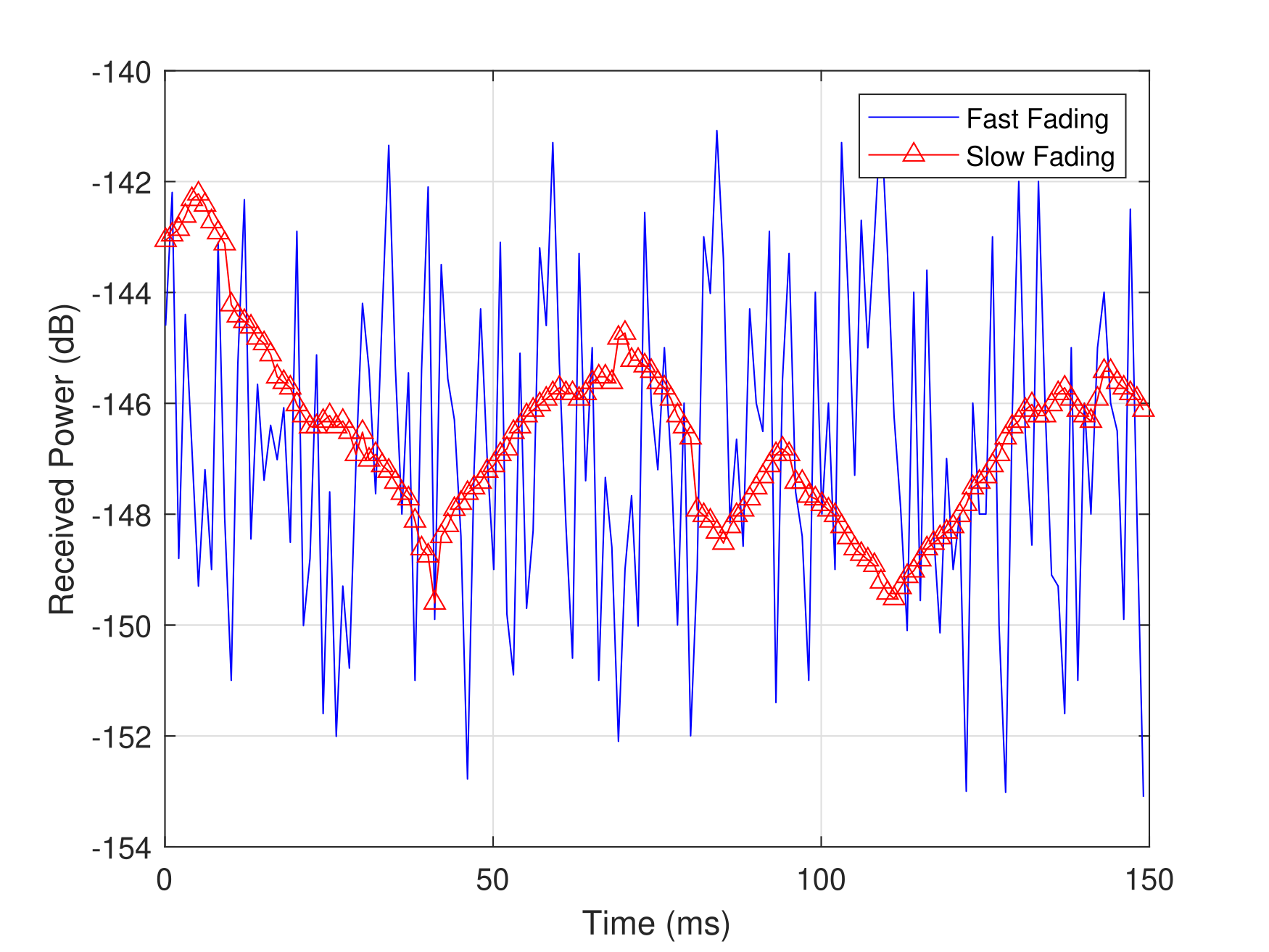}
\footnotesize
\caption{Power received in dB by the protein under fast and slow fading conditions.}
\label{fig:power_received}
\end{figure}

Fig.~\ref{fig:power_received} shows the received power in dB versus time  for a single EM signal realization under fast and slow fading scenarios. These scenarios are achieved by varying the velocity of RBCs. In fact, every time the nanoantenna transmits an EM signal, the strength of the impinging electric field is computed, and accordingly, the received power is calculated using~\eqref{eq:relationship}. 

\begin{figure}[h!]
 \centering
 \subfigure[]{%
  \includegraphics[height=4.9 cm, width=7 cm]{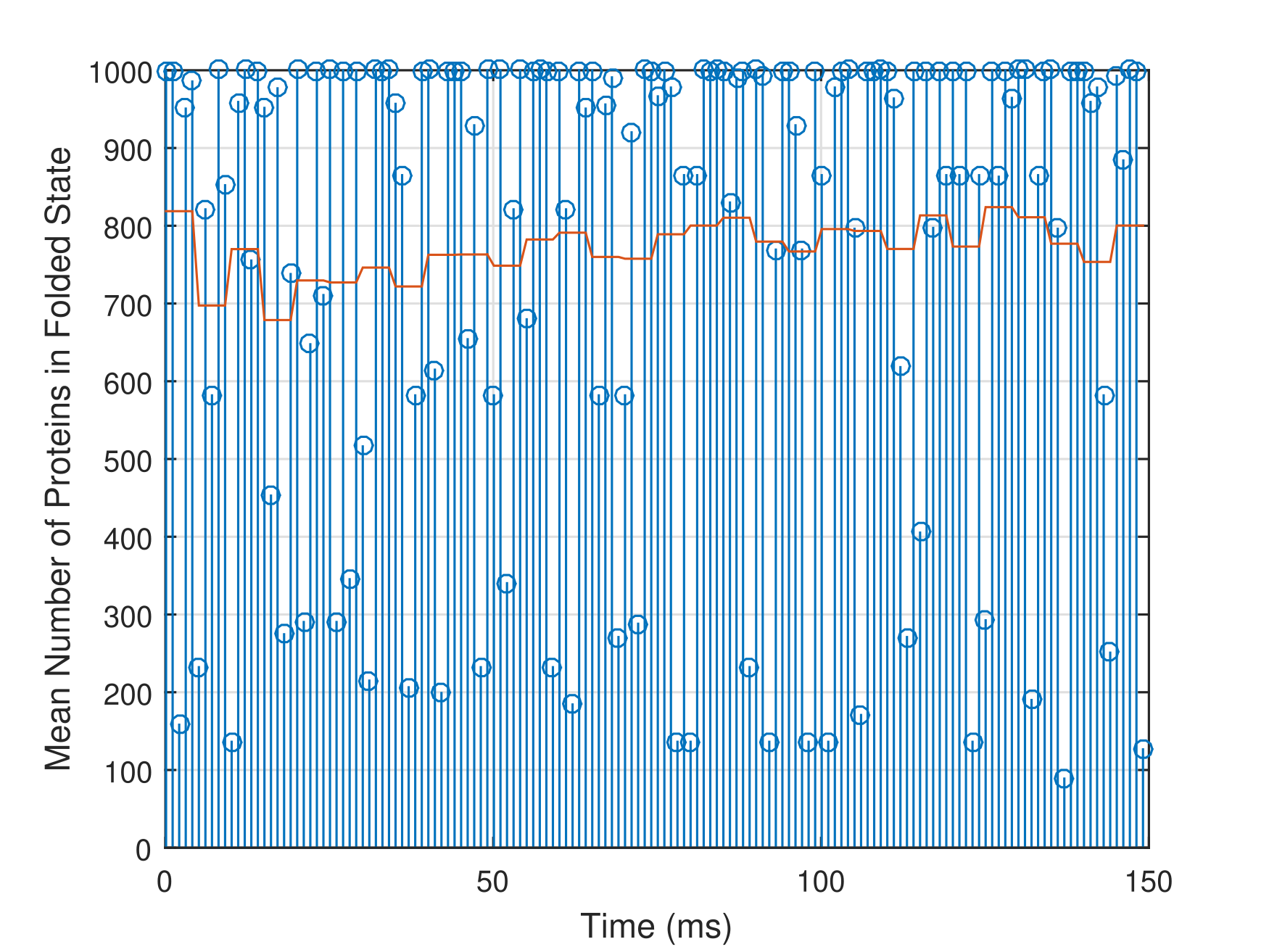}%
}

\subfigure[]{%
  \includegraphics[height=4.9 cm, width=7 cm]{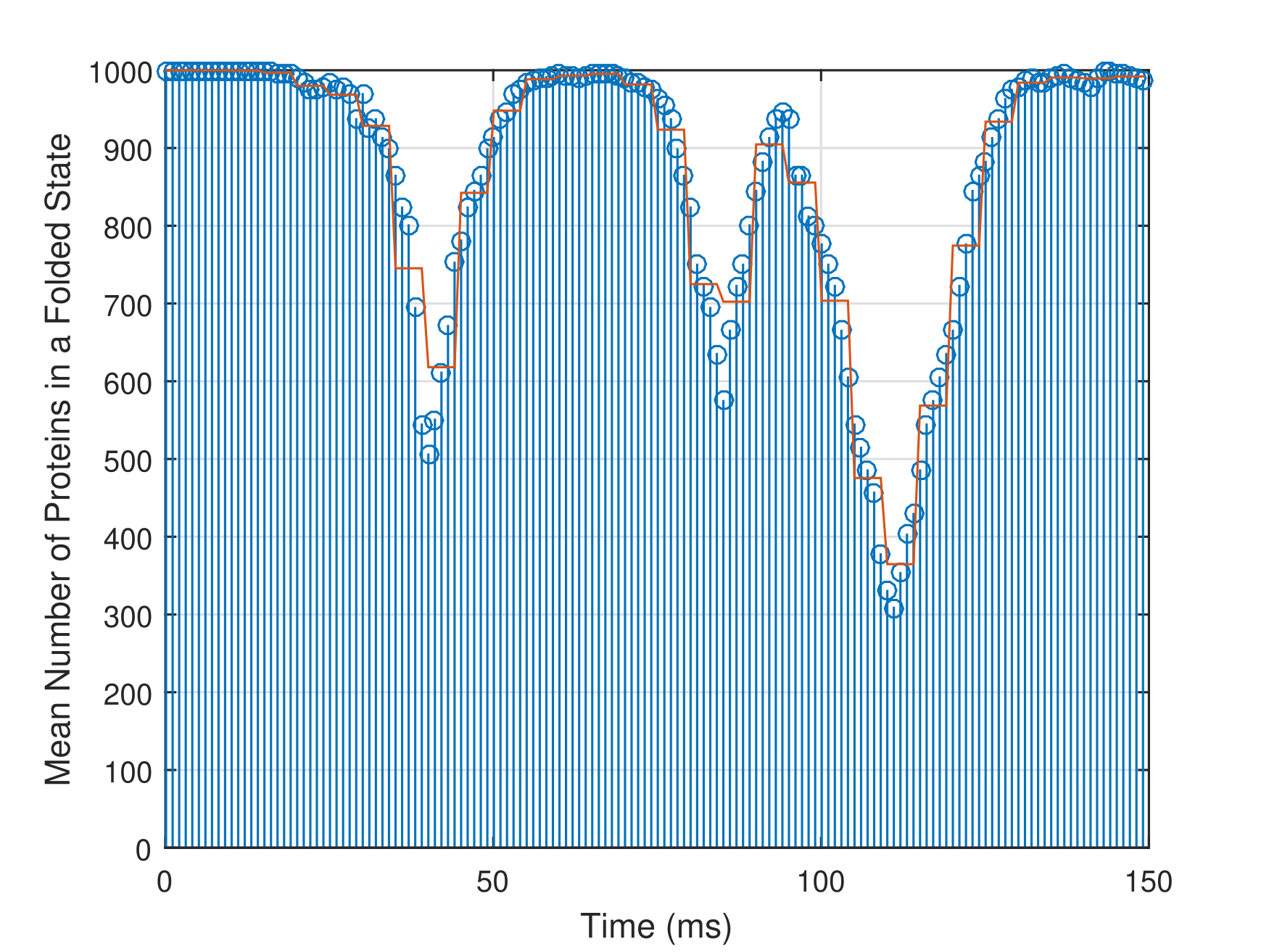}%
}
\caption{Mean number of proteins in a folded state (a)~under fast fading scenario. (b)~under slow fading scenario.}
\label{fig:fading}
\end{figure}

In addition, Figs.~\ref{fig:fading}(a) and~\ref{fig:fading}(b) illustrate the mean  number of folded proteins ($\mu=np_{F}$) versus time for the fast and slow fading scenarios, where $n=1000$. Under fast fading, we can notice from Fig.~\ref{fig:fading}(a) the impact of the fluctuating channel response on the protein folding probability. Since fast fading is an ergodic process,  we are interested in the time-average value, represented by the orange line in Fig.~\ref{fig:fading}(a). Consequently, from Fig.~\ref{fig:fading}(a), the average number of proteins in the folded state is $\mu=np_F=772$.

In Fig.~\ref{fig:fading}(b), we observe the effect of slow fading, characterized by a roughly constant amplitude and phase on the number of proteins in the folded state. In this case, the number of folded proteins averaged over time is $\mu=nE[p_F]=848$. The lower number of folded proteins observed under fast fading conditions, compared to the slow fading case, can be attributed biologically to the velocity of RBCs. The high velocity limits the cellular volume accessible to a protein polypeptide chain, thereby affecting its ability to attain a folded state. While one might argue that the velocity of RBCs in the fast fading example is high compared to real-world scenarios, different combinations of RBC and protein timescales can lead to fast-fading scenarios. Therefore, the provided example offers a clear visualization of its impact.

While the mean number of folded proteins is higher in the slow fading case, there is a noticeable drop in the number of folded proteins between $110$ and $150$~ms. This decline can be attributed to the fact that slow fading may affect the protein's configuration, causing it to become trapped in an intermediate state within a restricted region of phase space. In other words, the protein may become stuck in an intermediate state, preventing it from achieving a fully folded conformation. In such a scenario, the system would require the nanoantenna to re-transmit the signal, resulting in uncontrolled delays and stringent power requirements on the link between the nanoantenna and the targeted protein population.

\subsection{Selectivity under Fading Scenarios}

\begin{figure} [h!]
 \centering
\subfigure[]{%
  \includegraphics[height=5 cm, width= 7 cm]{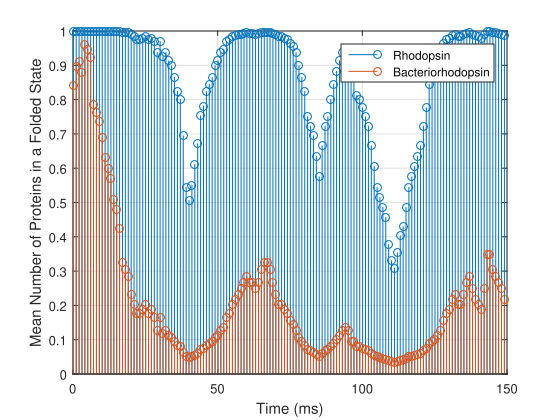}%
}
\subfigure[]{%
  \includegraphics[height=5 cm, width=7 cm]{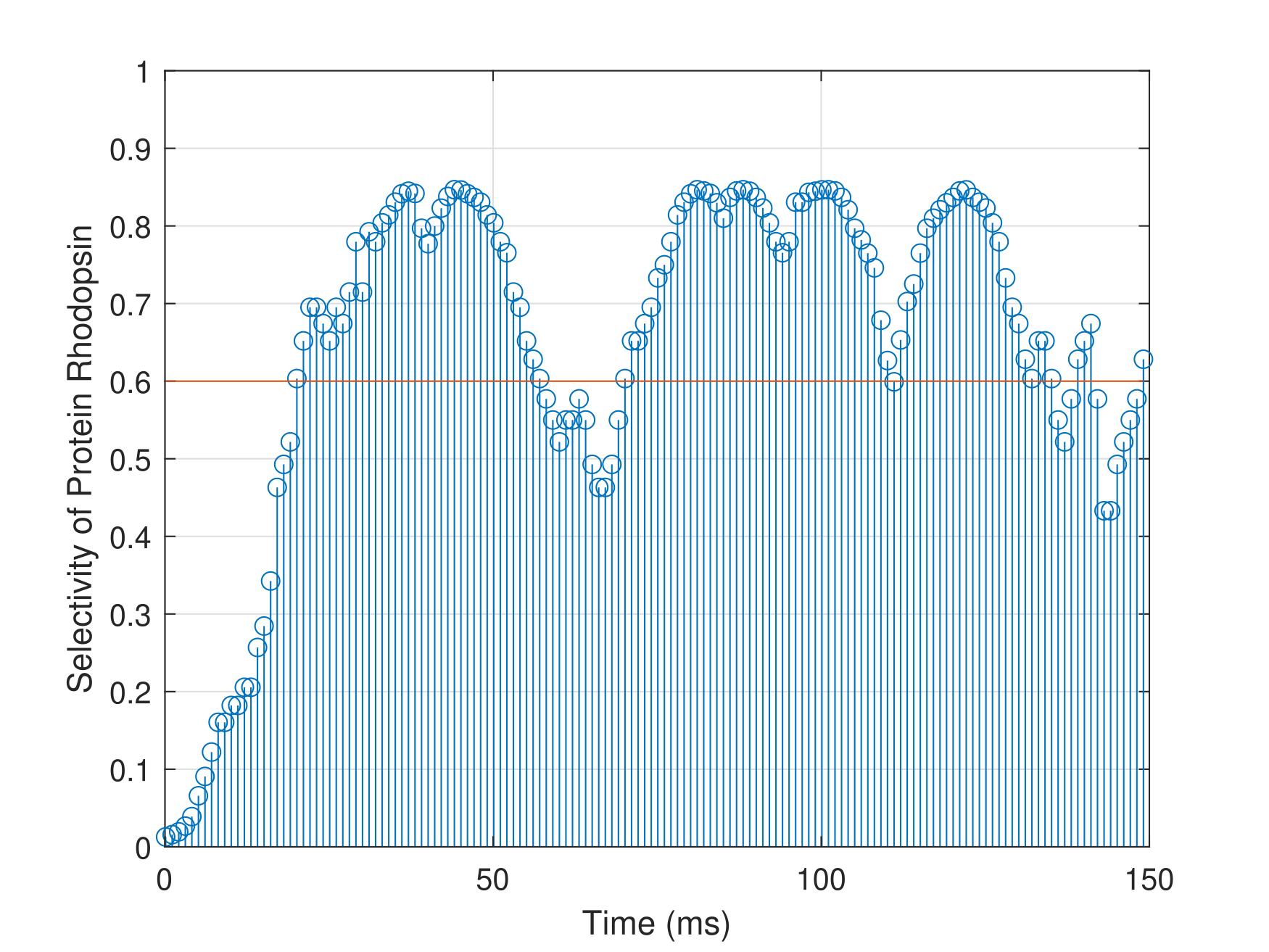}%
}
\caption{(a) Mean number of folded proteins of rhodopsin (targeted)
and bacteriorhodopsin. (b) Selectivity of rhodopsin ($\gamma_o$= 0.6).}
\label{fig:fading_selectivity}
\end{figure}

To further investigate this phenomenon, we examine the impact of slow fading on the system's selectivity. Therefore, we introduce another protein population, bacteriorhodopsin, in the vicinity of the targeted rhodopsin population to simulate a realistic protein network. The vibrational frequency of bacteriorhodopsin and its mechanical properties can be found in Table~\ref{table:parameters}. To calculate the selectivity, we use~\eqref{eq:metric2}, where we initially compare the mean number of folded proteins under slow fading conditions for both populations within the network, as illustrated in Fig.~\ref{fig:fading_selectivity}(a). It's worth noting that the nanoantenna frequency is tuned to the vibrational frequency of the rhodopsin (targeted) protein population.

Fig.~\ref{fig:fading_selectivity}(b) demonstrates the  selectivity of protein rhodopsin, with the  orange line representing the threshold we have set for  this example, $\gamma_o=0.6$. As expected, we can clearly observe a correlation between the performance in Fig.~\ref{fig:fading_selectivity}(a) and Fig.~\ref{fig:fading_selectivity}(b). In cases where the mean number of folded proteins of both populations is comparable, the selectivity is low,  while in cases where the the mean number of folded proteins in both populations differs significantly, the selectivity is high.

\begin{figure}[h!]
\centering
\includegraphics[width=0.4\textwidth]{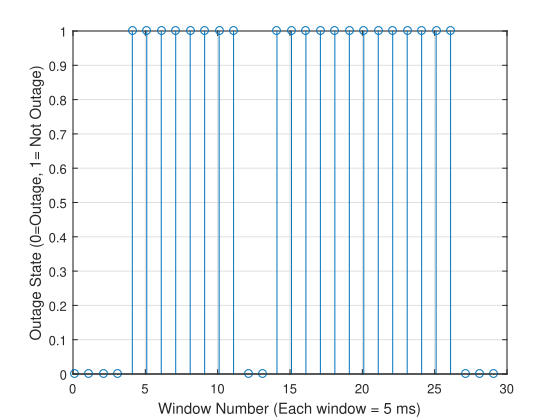}
\footnotesize
\caption{Outage versus window number when $\gamma_o=0.6$.
A value of zero indicates an outage, while a value of 1
indicates no outage.}
\label{fig:outage_window}
\end{figure}

Fig.~\ref{fig:outage_window} displays the outage status, i.e. whether we have an outage in the selectivity results provided in Fig.~\ref{fig:fading_selectivity}(b) or not, versus the window number.  To create Fig.~\ref{fig:outage_window}, we averaged the selectivity values over windows of size $T_s=5$ ms. Since our simulations span $150$~ms, this results in $30$ windows. We then determined, based on the assigned threshold, whether each window experienced an outage by checking if the selectivity value fell below the threshold.

For $\gamma_o=0.6$, we observed that out of the $30$ windows, $9$ windows experienced an outage. This results in a selectivity outage probability of $0.3$, indicating a $30\%$ chance that the system cannot differentiate between the rhodopsin and bacteriorhodopsin populations. Essentially, there exists a trade-off between the selectivity threshold set for the system and the potential for outages. A higher threshold leads to higher selectivity values but increases the likelihood of system outages, which imposes additional power constraints.

\begin{table}
\caption{Outage state vs Window Number}\begin{center}
\centering
\small
\label{table:threshold}
\begin{tabular}{|p{35pt}|p{75pt}|p{85pt}|}\hline
\begin{footnotesize}\textbf{Threshold} \end{footnotesize}&\begin{footnotesize} \textbf{Number of Windows with Outage} \end{footnotesize}&\begin{footnotesize} \textbf{Probability of Selectivity Outage} \end{footnotesize} \\\hline
 \begin{footnotesize}0.2  \end{footnotesize}&\begin{footnotesize} 2 \end{footnotesize} &\begin{footnotesize} 0.067\end{footnotesize}\\\hline
 \begin{footnotesize}0.4 \end{footnotesize}& \begin{footnotesize} 3  \end{footnotesize}&\begin{footnotesize} 0.1 \end{footnotesize}\\\hline
\begin{footnotesize}0.6 \end{footnotesize}& \begin{footnotesize}  9 \end{footnotesize}&\begin{footnotesize} 0.3\end{footnotesize}\\\hline
 \begin{footnotesize}0.8 \end{footnotesize}& \begin{footnotesize}  20 \end{footnotesize}&\begin{footnotesize} 0.67 \end{footnotesize}\\\hline
\end{tabular}
\end{center}
\end{table}

\subsection{Selectivity Optimization Results}
Experimental design, careful optimization, and validation are essential to ensure selectivity and minimize off-target effects. In fact, to design a system with maximum selectivity, we must determine the optimal values for both the nanoantenna transmit power and the operating frequency, as these are the controllable parameters. Therefore, we formulate a joint optimization problem for these parameters, as explained in Sec.~\ref{Sec:Sec5}, for both fast and slow fading scenarios.

\subsubsection{Fast Fading}
 Fig.~\ref{fig:selectivity_fading}(a) provides a pseudo-color plot for the selectivity  of the targeted rhodopsin population  in the presence of that of bacteriorhodopsin. The plot discriminates regions that should be targeted (given in yellow) from those that should be avoided (given in blue) in order to activate the correct protein population. From Fig.~\ref{fig:selectivity_fading}(a), we conclude that by tuning the nanoantenna to the resonant frequency of rhodopsin, i.e. $1.36$~THz, and to a Tx power of $30~\mu$W, maximum selectivity is achieved.

\subsubsection{Slow Fading}
\begin{figure}
 \centering
\subfigure[]{%
 \includegraphics[height=5.2 cm, width=7cm]{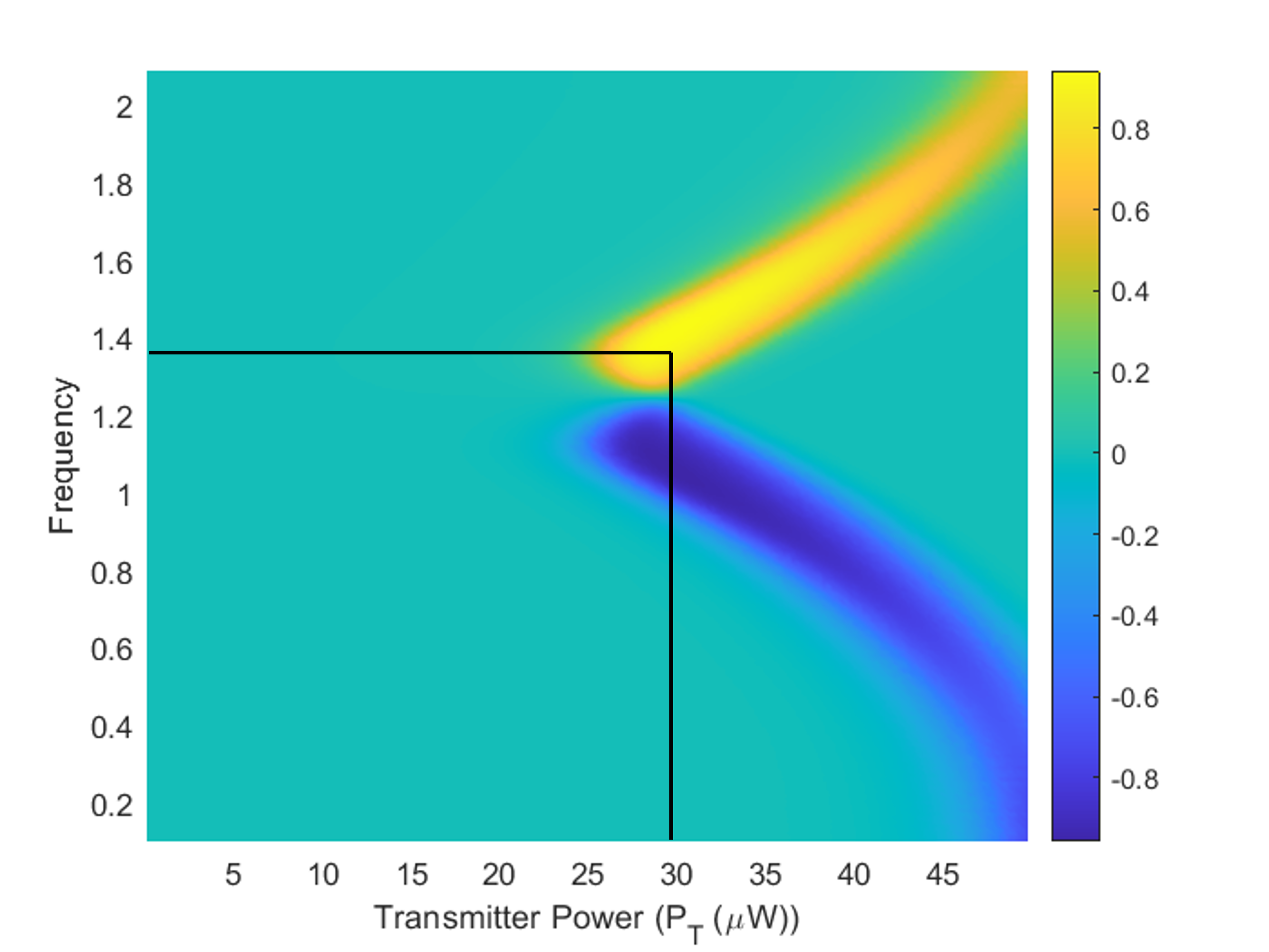}%
}
\subfigure[]{%
  \includegraphics[height=5.2 cm, width=7cm]{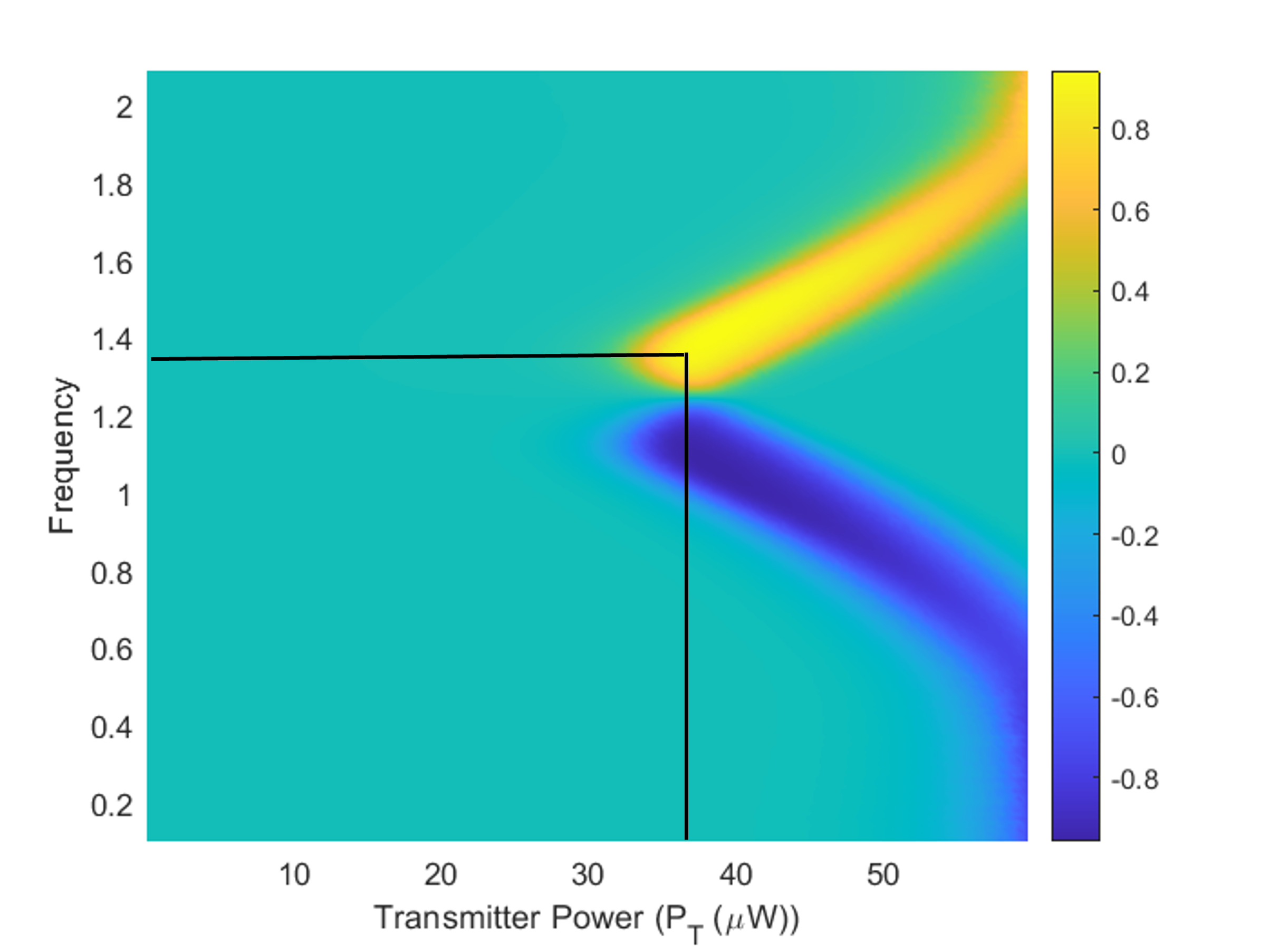}%
}
\caption{(a) Selectivity of rhodopsin (a)~under fast fading scenario. (b)~under slow fading scenario.}
\label{fig:selectivity_fading}
\end{figure}

Moving to the slow fading scenario, Fig.~\ref{fig:selectivity_fading}(b) demonstrates a pseudo-color plot for the selectivity  of the targeted rhodopsin population. We notice here that in comparison to Fig.~\ref{fig:selectivity_fading}(a), the system needs higher Tx power, around $38~\mu$W, to achieve a maximum selectivity score. This power is required to overcome the possible outages in the system that may occur in occasions of deep fades. Through the presented results, one can determine the amount of power that will prevent outages and concurrently  prevent the activation of the untargeted protein population. Hence, when designing biosensors for targeted therapy applications, engineers must carefully take into account all these tuning parameters.   

It should be noted that operating nanonetworks at power levels in the microwatt range enables the creation of energy-efficient, miniaturized devices with improved heat dissipation capabilities. This approach aligns with the unique requirements and constraints associated with nanoscale communication and integration, making it feasible to deploy nanonetworks in various applications, ranging from healthcare and environmental monitoring to industrial and scientific domains.

\section{Conclusions} 
\label{Sec:Sec7}
In this work, we analyze the influence of fading on THz-induced protein interactions. We investigate how the attenuation in the transmitted THz signal affects the ability to achieve selective protein folding behaviors in the desired protein population. Under fast fading channel conditions, we observed that the protein folding probability is lower than in the case of slow fading, primarily due to the variability in the channel response

While it's easier to track the changes imposed by the channel under slow fading conditions, the system may still experience outages, resulting in significant delays in receiving the nanoantenna signal and placing additional demands on the power requirements of the system.

Through the examination of various fading scenarios achieved by varying the velocity of RBCs, we illustrate the influence of the cellular environment on protein dynamics. This analysis can provide insights into the effects of molecular crowding and hindered diffusion on protein folding. By taking into account channel variability, we can design biosensors and nanoantennas that align with medical requirements and fully harness the potential of the intra-body communication channel.

However, one of the current limitations of the developed model is the absence of experimental validation. While theoretically, THz antennas could be miniaturized due to the short wavelength of THz-EM waves, current systems often rely on larger and more obtrusive equipment, such as tabletop microscopes and spectroscopy systems. These factors create barriers that constrain the potential and possibilities of this emerging field. Research efforts are therefore needed to address these challenges and enhance the feasibility of intra-body THz experiments for clinical applications. Advancements in nanotechnology offer a promising solution, as they provide diverse opportunities to enhance the performance, sensitivity, and applicability of THz experiments across various fields.

\bibliographystyle{IEEEtran}
\bibliography{IEEEabrv,./references}

\end{document}